\documentclass[prd,twocolumn,showpacs,linenumbers,groupedaddress,superscriptaddress,amsmath,amssymb]{revtex4}

\usepackage{graphicx}
\usepackage{dcolumn}
\usepackage{bm}
\usepackage{amssymb}
\usepackage{enumerate}
\usepackage{lineno}
\usepackage{multirow} 
\usepackage{float}

\newcommand\g{\gamma}

\newcommand\emu{e^- Z \to e^- Z \g; \g \to \mu^+ \mu^-}

\newcommand\ainv{\varphi  \to invisible}

\def\address{\@ifstar{\address@star}%
  {\@ifnextchar[{\address@optarg}{\address@noptarg}}}

\usepackage[printwatermark]{xwatermark}
\usepackage{xcolor}
\usepackage{graphicx}
\usepackage{lipsum}

\begin{document}


\title{ Leptonic scalar portal: Origin of  muon $g-2$ anomaly and  dark matter? }


\author{S.~N.~Gninenko}
\affiliation{\it Institute for Nuclear Research, 117312 Moscow, Russia}
\author{ N.~V.~Krasnikov}\affiliation{\it Institute for Nuclear Research, 117312 Moscow, Russia}\affiliation{\it  Joint Institute for Nuclear Research, 141980 Dubna, Russia}

\date{\today}
\begin{abstract}
We present a model explaining both  the  4.2 $\sigma$ muon $g-2$ anomaly and  the relic density of dark matter (DM)  in which  DM interacts with the Standard Model (SM) via a scalar portal boson $\varphi$ carrying both dark and SM leptonic numbers, and mediating a nondiagonal interaction between the electron and muon that allows  $e \leftrightarrow \mu$ transitions. The $\varphi$ could be produced in  high-energy electron scattering off a target nuclei in 
the reaction $e  Z \to \mu Z\varphi$ followed by the prompt invisible decay $\varphi~\to$~DM particles and searched for in events with  large missing energy accompanied by a single outgoing muon   in the final state.  Interestingly, several events with a similar signature have been observed  in a data sample of $\simeq 3\times 10^{11}$ electrons on target collected during 2016-2018 for the search for light dark matter in the NA64 experiment at the CERN SPS [PRL {\bf 123}, 121801 (2019)]. Attributing so far these events to background  allows us to set first constraints on the $\varphi$  mass and couplings while  leaving  at the same time decisively probing the origin of these  events and a large fraction of the remaining parameter space to a near exiting future with the upgraded  NA64 detector or other planned experiments.  
\end{abstract}
\maketitle
\par The recent precise determination  of the  anomalous magnetic 
moment of the positive muon $a _{\mu}= (g-2)_\mu/2$ from the  experiment E989 at FNAL  \cite{fnal} confirmed the previous measurements of Ref.\cite{bnl}, and 
 gives  result which is about $4.2 \sigma$ higher than the Standard Model (SM) prediction, see, e.g.,  \cite{th1,th2,th3,th4,th5,th6,th7,thh8,thh9,thh10,thh11,thh12}
\begin{equation}
a_{\mu}^{exp} - a_{\mu}^{SM}  =  (251 \pm 59) \times 10^{-11}
\label{eq:g-2} 
\end{equation}
This result may signal the existence of new physics (NP) below the electroweak scale ($\ll 100$ GeV), see e.g., Ref.\cite{g-2_np}.
For example, one of the most attractive 
  explanations of the anomaly suggests the existence of a sub-GeV gauge boson, which can be probed in a near future at a fixed-target 
  experiment, see e.g. \cite{gk1,gkm,gk2,laura, laurakirp, Kirpichnikov:2020tcf, na64epjc21, krnjaic, krnjaic1, krnjaic2, chen}. 
\par Another motivation for searches of NP in the low-mass range  come from the  dark matter (DM)  sector.
Despite many intensive searches at the accelerator and in nonaccelerator experiments, still little is known about the origin and dynamics of the  dark sector itself. 
One difficulty so far is that DM   can be probed only through its gravitational interaction. Thus, sensitive  searches for possible portals that could transmit new feeble  interaction between the ordinary and dark matter are crucial and, indeed, they have  received significant attention in recent
 years  \cite{jr,Essig:2013lka,report1,report2,pbc-bsm}.  
\par The  goal of this work is to show that  both  the  4.2 $\sigma$ muon $g-2$ anomaly and  the relic density of dark matter (DM) could be explained by a model 
 in which  DM interacts with the Standard Model (SM) via a scalar portal boson $\varphi$ carrying  SM $L_e$ and $L_\mu$ leptonic numbers. The $\varphi$ mediates a nondiagonal interaction between the electron and muon that allows  $e \leftrightarrow \mu$ transitions, while the leptonic numbers are conserved.
   Similar models were considered in the recent past, but unlike the present model they 
considered diagonal interactions transmitted by a mediator carrying  different quantum numbers, 
see, e.g. Ref. \cite{dama,cdmz,bbat,gk161,gk162}.   It is assumed that the  $\varphi$ decays predominantly invisibly, $\Gamma (\varphi \rightarrow invisible)/\Gamma_{tot} \simeq 1$, e.g., into dark sector particles, thus escaping  stringent constraints placed today on the visible   decay modes of the $\varphi$ into SM particles from collider, fixed-target, and atomic experiments  \cite{pdg}.  
  The most stringent limits on the invisible  $\varphi$ in the sub-GeV mass range are obtained, so far, for the case of scalars $\varphi$  coupled to electron and muon by the low-energy experiments searching for the muon decay  $\mu \to e \varphi$ \cite{pdg}, leaving a large area of the parameter space for the leptonic $\varphi$  still unexplored.
Therefore in the following  we assume that $m_\varphi \gtrsim  m_{\mu}$.  
 \begin{figure}[tbh!!]
\includegraphics[width=.2\textwidth]{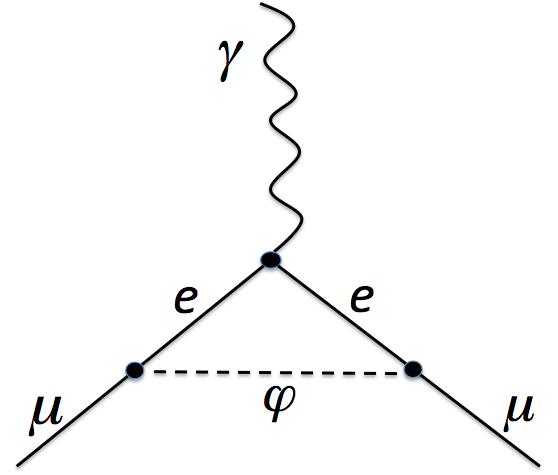}%
\vskip-0.cm{\caption{One-loop contribution of the leptonic scalar $\varphi$ to $\Delta a_\mu$. \label{fig:diagr}}}
\label{fig:g-2} 
\end{figure}
\par Consider  the interaction of a complex scalar mediator $\varphi(x)$ with electrons and muons, namely 
\begin{equation}
L_{\varphi \mu e}  = -h_{\mu e}\bar{e}_L\mu_R  \varphi   + H.c. \,,
\label{Lmue}
\end{equation}
where $e_L = (\frac{1 - \gamma_5}{2})e $, $\mu_R = (\frac{1 + \gamma_5}{2})\mu$,
The interaction (\ref{Lmue}) is invariant under the 
$L_e$, $L_{\mu}$ flavor 
global transformations $\varphi(x) \rightarrow exp(i\alpha_e + i\alpha_{\mu}) \varphi(x),~ \mu(x) \rightarrow exp(-i\alpha_{\mu}) \mu(x)$, and $e(x) \rightarrow exp(i\alpha_{e}) e(x)$.
Due to the postulated  global symmetry,  the Lagrangian (2) contains only nondiagonal terms like 
$-h_{\mu e}\bar{e}_L\mu_R$,  and flavor diagonal terms  $-h_{ee}\bar{e}e\varphi$ and  $-h_{\mu\mu}\bar{\mu}\mu \varphi$  are prohibited.
As a consequence,  for massless neutrino the interaction (\ref{Lmue}) transmitted by the leptonic $\varphi$  conserves both muon and electron 
lepton numbers. The interaction  (\ref{Lmue}) leads to additional contributions to the electron and 
muon $(g-2)$.  One-loop contribution to $a_\mu$ is shown in  Fig.\ref{fig:g-2} 
and it  reads \cite{muonmoment}
\begin{equation}
\Delta a_{\mu} = \frac{h^2_{\mu e}}{16\pi^2}\frac{m_{\mu}^2}{m^2_\varphi}L \,,
\label{mu1}
\end{equation}
\begin{equation}
L = \frac{1}{2}\int^1_0 dx \frac{2x^2(1-x)}{(1-x)(1 -\lambda^2 x) + (\epsilon\lambda)^2 x} \,,
\label{mu2}
\end{equation}
where $\epsilon = \frac{m_e}{m_{\mu}}$ and $\lambda = \frac{m_{\mu}}{m_\varphi}$.
For electron magnetic magnetic moment we must replace $m_{\mu}$ to $m_e$ and $m_e$ to $m_{\mu}$ 
in formulas \eqref{mu1}, \eqref{mu2}. 
For $m_\varphi \gg m_{\mu}$ one can find \cite{muonmoment} that
\begin{equation}
\Delta a_{e({\mu})} = \frac{h^2_{\mu e}}{48\pi^2} \frac{m^2_{e({\mu})}}{m^2_\varphi} \,,
\label{aemu}
\end{equation}
and $\frac{\Delta a_e}{\Delta a_{\mu}} = (\frac{m_e}{m_{\mu}})^2$.
If we assume that the additional interaction explains the muon anomaly  (\ref{eq:g-2}), then
\begin{equation}
h_{\mu e} = (1.1 \pm 0.1) \times  10^{-3} (\frac{m_\varphi}{m_{\mu}}) \,.
\label{hmuemuon}
\end{equation} 
for $m_\varphi \gg m_{\mu}$.
As it was mentioned previuosly in the rest of the  paper we assume 
that $m_\varphi > m_{\mu} $. This assumption allows us to 
prohibit the decay $\mu \rightarrow e \varphi $ for which experimental data restrict rather strongly 
the coupling constant $h_{\mu e}$. 
For our estimates we shall use the conventional  point  $m_\varphi = 3 m_{\mu}$ resulting in 
\begin{equation}
h_{\mu e} = (3.3 \pm 0.3) \times 10^{-3} \,.
\label{reper}
\end{equation}
for explaining the value (\ref{eq:g-2}).

The $SU_L(2) \otimes U(1)$ invariant generalization of the interaction  (\ref{Lmue}) is 
\begin{equation}
L_{\mu e, gen} = -\frac{h_1h_2}{M} (\bar{\nu}_e, \bar{e})_{L}H \varphi \mu_{R} + H.c. \,,
\label{mu e,gen}
\end{equation}
where $\frac{h_1h_2<H>}{M}=h_{\mu e}$, and 
  $<H> = 174~GeV$ is the vacuum expectation value of the Higgs isodoublet $H$. In the unitary gauge 
 $H = (0,\frac{h}{\sqrt{2}} + <H>)  $, where $h$ is the Higgs field. Note that the complex scalar mediator $\varphi(x)$ is a singlet
under the $SU_c(3)\otimes SU_L(2)\otimes U(1)$  SM gauge group. Due to possible interaction $L = -\lambda_{H\phi} H^+H\varphi^*\varphi$ of 
the scalar $\varphi$ with the Higgs isodoublet,   Higgs boson would decay invisibly into a $\varphi$ pair,  $h \rightarrow \varphi\varphi^*$, with a rate given by 
$\Gamma(h \rightarrow \varphi\varphi^*) = \frac{\lambda^2_{H\varphi}v^2}{16\pi m_h}(1- \frac{4m^2_\varphi}{m^2_h})^{1/2}$ assuming that the invisible decay $\varphi \to DM~particles$ is dominant (see below).  Here $m_h$ is the Higgs boson mass and $v = 246~GeV$. From the existing  bounds on the Higgs boson invisible decay width \cite{pdg} one can obtain an  upper bound on the coupling constant $\lambda_{H\varphi}  \leq 0.01 $.
 The interaction (\ref{mu e,gen}) is nonrenormalizable and it conserves both $L_e$ and $L_{\mu}$ flavor numbers in the 
approximation of massless neutrino.
One can obtain the effective nonrenormalizable interaction (\ref{mu e,gen}) from the renormalizable interaction with vectorlike fermion $E$, namely 
\begin{equation}
L_{E_R \mu e} = -(h_1(\bar{\nu}_e, \bar{e})_LHE_R  + h_2\bar{E}_L\mu_R \varphi  + H.c.) - M\bar{E}E 
\end{equation}
\par Suppose the $\varphi$-boson interacts with dark mater particles. Several models can 
be considered. 
First, the $\varphi(x)$ field could have interaction with two dark matter complex  scalars   $s_1(x)$ and $s_2(x)$  given by 
\begin{equation}
L_{\varphi s_1s_2} = g_{\varphi s_1s_2}\varphi s_1s_2 + H.c.
\label{aa_1a_2}
\end{equation}
Note that the coupling constant  $g_{\varphi s_1s_2}$ has the dimension of the mass.
The interaction (\ref{aa_1a_2}) is invariant under global transformations
$\varphi  \to exp(i\alpha_1 + i\alpha_2))\varphi $,  and $s_i \to exp(-i\alpha_i)s_i$, with $i=1,2$.
As a consequence  in the approximation of massless neutrino both $L_e$ and $L_{\mu}$ lepton 
flavors are conserved.
\par Consider another model, when the  scalar $\varphi$ interacts with two light dark matter 
fermions $\psi_1$ and $\psi_2$  with the Lagrangian      
\begin{equation}
L_{\varphi \psi_1\psi_2} = g_{\varphi \psi_1\psi_2}\varphi     \bar{\psi}_1\psi_2 + H.c.
\label{apsi}
\end{equation}
Again interaction (\ref{apsi}) conserves both $L_e$ and $L_{\mu}$ lepton flavors. 
For the Lagrangians (\ref{aa_1a_2}) and  (\ref{apsi}) the $\varphi$ decay rate 
into   $s_1, s_2$ and  $\psi_1, \psi_2$ DM particles  is 
\begin{equation}
\Gamma(\varphi  \rightarrow s_1 s_2) = \frac{g^2_{\varphi s_1s_2}}{8\pi}\frac{p_1}{m^2_\varphi}, 
\end{equation}
and
\begin{equation}
\Gamma(\varphi  \rightarrow \psi_1 \psi_2) = \frac{g^2_{\varphi \psi_1\psi_2} p_1}{4\pi}
(1 - \frac{(m_1 +m_2)^2}{m^2_\varphi})
\end{equation}
respectively, and $p_1=\frac{[(m^2_\varphi -(m_{1}+m_2)^2)(m^2_\varphi -(m_1-m_2)^2)]^{1/2}}{2m_\varphi}$ is the momentum of the particle $1$ in 
the rest frame of the $\varphi$,  and $m_1$ and $m_2$ are the masses of particles $1$ and 
 $2$.
Here, we assume that $ m_\varphi > m_1 + m_2$.
  The decay width of $\varphi$ into $\mu^+ e^-$ is given by 
\begin{equation}
\Gamma(\varphi  \rightarrow \mu^+ e^{- }) = \frac{h^2_{\mu e}p_e}{8\pi}
(1 - \frac{m_{e}^2 +m_{\mu}^2}{m^2_\varphi}), 
\end{equation}
where $p_e$
is the electron momentum in the center of mass frame. 
\begin{figure*}[tbh!!]
\includegraphics[width=.7\textwidth]{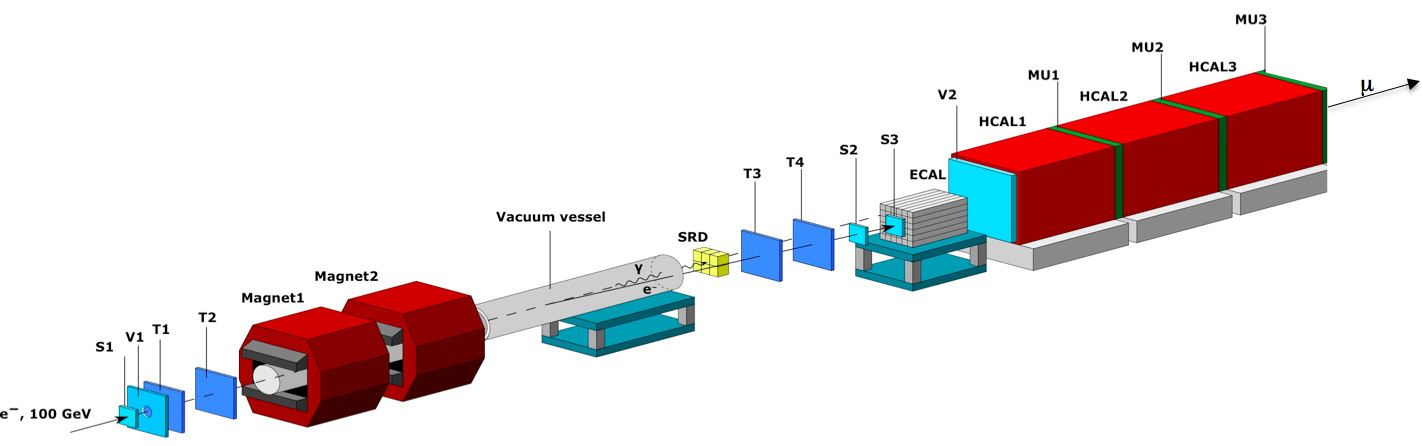}%
\vskip-0.cm{\caption{Schematic illustration of the setup to search for the  $\ainv$  decay of the leptonic scalar $\varphi$
produced in the reaction  $eZ\to  \mu Z \varphi $ of 100 GeV $e^-$'s incident  on the active ECAL target of the NA64e experiment.\label{setup}}}
\end{figure*} 
The annihilation cross sections of $s_1$, $s_2$ 
into $\mu e$ pair in the nonrelativistic approximation in
$s$-wave is   
\begin{eqnarray}
\sigma_{an}(s_1s_2 \rightarrow e^{+}\mu^{-})v_{rel} 
= \sigma_{an}(s_1s_2 \rightarrow e^{-}\mu^{+})v_{rel} \nonumber \\
  =  |M|^2_1  \frac{ p_{ecm}  }{16\pi(m_1 + m_2)m_1m_2} 
\label{annihilation}
 \end{eqnarray}
where
\begin{equation}
|M|^2_{1} =  
g^2_{\varphi s_1s_2} h^2_{\mu e} \frac{ ((m_1 +m_2)^2  - m^2_e - m^2_{\mu})}{(m^2_\varphi - (m_1 + m_2)^2)^2}
\end{equation}
and $p_{ecm}$ is the momentum of electron in the center of mass frame\footnote{
$p_{ecm} = \sqrt{E^2_{ecm} -m^2_e}$, $E_{ecm} = \frac{s + m^2_e - m^2_{\mu}}{2\sqrt{s} } $, 
$s = (m_1 + m_2)^2$.}
For the simplest  case of dark matter particles with equal masses,  $m_1 = m_2 \gg m_{\mu}$,  the annihilation cross section is 
\begin{equation}
\sigma_{an}(s_1s_2 \rightarrow e^{+}\mu^{-})v_{rel} = \frac{h^2_{\mu e}g^2_{\varphi s_1s_2}}
{8\pi(m^2_\varphi - 4 m^2_1)^2} 
\end{equation}
The treatment of  general case  with  nonequal masses  is  straightforward and does not qualitatively changes our main conclusions. 
For  fermions $\psi_1$, $\psi_2$ in the nonrelativistic approximation 
the annihilation cross section  $\sigma_{an}(\psi_1\psi_2 \rightarrow e^{-} \mu^{+} )v_{rel}$  
is given by the formula (\ref{annihilation}) where 
\begin{eqnarray}
|M|^2_2 =      \frac{g^2_{\varphi \psi_1\psi_2}}{2} h^2_{\mu e} \frac{ 1 }{(m^2_\varphi - (m_1 + m_2)^2)^2} \nonumber \\
\times ((m_1 + m_2)^2 - m^2_e - m^2_{\mu}) m_1m_2 v^2_{rel}
\end{eqnarray}
The total annihilation cross section is given by  
\begin{equation}
\sigma_{an,tot}v_{rel} = \frac{  h^2_{\mu e}   g^2_{\varphi \psi_1\psi_2}m^2_1v^2_{rel}}{8\pi (m^2_\varphi - 4m_1 ^2)^2} \,,      
\label{pwavean}
\end{equation}
where $\sigma_{an,tot} =  \sigma_{an}(\psi_1\psi_2 \rightarrow e^{-} \mu^{+} ) +  \sigma_{an}(\psi_1\psi_2 \rightarrow e^{+} \mu^{-} )$.    
Thus, we see that in the nonrelativistic limit model with scalar DM  particles has $s$-wave 
behavior that  contradicts to the Planck data \cite{Planck}\footnote{For $m_2 > m_1 +m_e + m_{\mu}$ 
the heaviest DM  particle $s_2$ is unstable and it decays into 
the lightest DM particle $s_1$ and $\mu e$ pair, namely 
$s_2 \rightarrow s^*_1 \mu^- e^+$ that in full analogy with the case of pseudo Dirac light matter 
allows us  to escape Planck restrictions.}
\par For the model with fermionic DM,  we have $p$-wave behavior for the annihilation cross section that allows us to escape 
Planck restrictions \cite{Planck}. 
 We assume that at the early Universe light DM is in 
equilibrium with ordinary matter. From the requirement that the relic density of  DM  
is explained by the model,  we can estimate the coupling constant $g_{\varphi \psi_1\psi_2}$ using standard 
formulae  for  calculations of the  DM density \cite{c1,c2,c3,c4,c5}. For this estimate we assume that  the p-wave annihilation cross section 
 $<\sigma_{an}v_{rel}> = O(1)~pb$,  and  the average relative velocity 
of annihilating DM particles $<v_{rel}> \sim c/3$ which corresponds to 
the observed DM density of the Universe  \cite{sp}. Consider the simplest example with 
$m_1 = m_2 \gg m_{\mu}$. As a consequence of the formula (\ref{pwavean}) we find that
\begin{equation}
 \frac{  h^2_{\mu e}   g^2_{\varphi \psi_1\psi_2}m^2_1}{4\pi (m^2_\varphi - (m_1 + m_2)^2)^2} 
= O(10~pb)
\end{equation}
For the case $m_\varphi = 3m_1$ we find 
\begin{equation}
  h_{\mu e}   g_{\varphi \psi_1\psi_2} \sim 10^{-3} (\frac{m_\varphi}{GeV})
\end{equation}
In the assumption that the model explains muon $g-2$  we find that
 $g_{\varphi \psi_1\psi_2} \sim 0.1  $ and it depends rather weakly on the $\varphi$  mass. 
 As a consequence we obtain that    $g_{\varphi \psi_1\psi_2} \geq h_{\mu e}$ for $m_\varphi \leq 10~GeV$  and the mediator $\varphi$ decays mainly invisibly into 
DM particles.
So we find  that our  model can  explain both the $(g-2)_\mu$ anomaly and the dark matter relic abundance.
\par Let us briefly discuss constraints on the model from the existing data. 
Note, that as  both $L_e$ and $L_\mu$  lepton  numbers are conserved the  muonium to antimuonium conversion, $\mu^+ e^- \to \mu^- e^+$, 
 is prohibited. As we already mentioned, assuming the invisible $\varphi$ boson decay is predominant, i.e. $\Gamma(\varphi \to all) \simeq \Gamma(\varphi \to DM)$, 
 the constraints on coupling $  h_{\mu e} $ from Higgs boson decays are quite modest. The interaction \eqref{Lmue} would also result in  LFV-like semivisible $Z$-boson decays
 $ Z \to  e^\pm e^{\mp} \to e^\pm \mu^\mp \varphi ;  \varphi  \to invisible$ and  $ Z \to \mu^\pm \mu{^\mp} \to \mu^\pm e^\mp \varphi ;  \varphi  \to invisible$.
For $m_\varphi \ll m_Z$ the branching ratio 
$\frac{\Gamma(Z \to\mu^{\pm} e^{\mp} \varphi )}{\Gamma(Z \to e^+e^-)}  \sim \frac{h^2_{\mu e}}{4\pi^2}$.
Assuming $m_\varphi = 3 m_{\mu}$ and   $h_{\mu e} = 3.3 \times 10^{-3}$, one gets 
$\frac{\Gamma(Z \to \mu^{\pm} e^{\mp}\varphi )}{\Gamma(Z\to all)} \sim 10^{-8} $. This can be compared  with 
the best  experimental constraint  $\frac{\Gamma(Z \to \mu^{\pm} e^{\mp}) }{\Gamma(Z\to all} < 7.5 \times 10^{-7}$ \cite{pdg} 
which is much weaker. Assuming that for the missing mass $\Delta m_{miss} \lesssim 5$ GeV, which is the experimental resolution of the $Z$-mass peak \cite{atlas},  the decays $Z\to e^\pm \mu^\mp \varphi $ and $Z\to e^\pm \mu^\mp$
are indistinguishable, one could get  $h_{\mu e} \lesssim   3\times 10^{-2}$  for the  sub-GeV $m_\varphi$  region . 
 Our model also predicts the $K \to \mu \nu \to  e \nu \varphi $ decay chain  with 
the branching ratio $Br(K\to e \nu \varphi ) \sim O(\frac{h^2_{\mu e}}{8 \pi^2}) \sim 2\times  10^{-7}$. 
By using the experimental constraints $Br(K\to e \nu \nu \overline{\nu}) < 6 \times 10^{-5}$  for  the momentum range  220-230 MeV/c 
\cite{k-e3nu} and a phase-space spectrum for the $K\to e \nu \varphi $ decay  
one can obtain modest bounds  $h_{\mu e} \lesssim 7\times 10^{-2}$ for the mass range $m_\varphi \lesssim 200$ MeV. 
For $ m_K - m_\varphi \gtrsim 250$ MeV bound  from $K \to e \nu \varphi $ decay does not work due to kinematics constraints of Ref.\cite{k-e3nu}.
\par The stronger limits on coupling $h_{\mu e}$ comes from anomalous magnetic moment of 
muon.  By using Eq.(\ref{eq:g-2})  we obtain that at $3\sigma$ level
the contribution of new physics to  $(g-2)_\mu$  is $\Delta a_{\mu} = a_{\mu}^{exp} - a_{\mu}^{SM}  \lesssim  428 \times 10^{-11}$.
Using Eqs.(\ref{mu1} -  \ref{aemu}) one gets for $m_\varphi > m_{\mu}$, that     $h_{\mu e} \lesssim 1.42 \times 10^{-3} (\frac{m_\varphi}{m_{\mu}})$ 
at $3\sigma$ level. For $m_\varphi = 3 m_{\mu}$ we find that $h_{\mu e} \leq 4.26 \times 10^{-3}$.
Note that bound from $\Delta a _{\mu} $ gets weaker proportionally  to $m_\varphi$,  and for large masses 
$m_\varphi \gtrsim$ a few  GeV the ATLAS  bound from the $Z$-decays  becomes stronger. 
\par Additional constraints can be obtained  from the NA64e experiment. For the sensitivity  estimate   
we will use NA64e  results on the search for light DM production in  invisible decays of dark-photon ($A'$) mediator obtained with  $n_{EOT}=2.84\times 10^{11}$ 100 GeV electrons on target (EOT)  \cite{na64prl17,na64prd18,na64prl19}.  If the  $\varphi$ exists, it could be  produced in the reaction 
\begin{equation}
eZ \to \mu Z \varphi ; \varphi \to invisible
\label{e-react}
\end{equation}
of high-energy electrons  scattering off nuclei of an active target of a hermetic NA64e detector, followed by the prompt invisible $\varphi$  decay   into DM  particles, which carry away part of the beam energy. 
A more detailed description of the NA64e detector can be found in Refs.\cite{na64prd18,na64prl19}. Below, its main relevant features  will be briefly mentioned. The detector
schematically shown in Fig.\ref{setup} employed  a  100 GeV pure electron beam, using the H4 beam-line of the CERN's North Area
with intensity of up to  $\simeq 10^7$ electrons per spill. The beam electrons impinging the target are measured by a 
magnetic spectrometer  consisting of two successive  dipole magnets  and a low-material-budget tracker chambers $T1-T4$ 
\cite{Banerjee:2015eno}. The beam electrons are  tagged by detecting  the synchrotron radiation (SR) emitted  by them in the magnets with the SRD counter \cite{na64srd}.  The active target  is  an electromagnetic calorimeter (ECAL), followed by  a hermetic hadronic calorimeter (HCAL) consisting of three consecutive modules. The HCAL  and the counters MU1-MU3, located between the modules,  are used as an efficient veto against hadronic secondaries  and also for identification  of muons produced in  the primary $e^-$ interactions  in the final state.
\par The signature of the reaction (\ref{e-react})  would be an event  with a fraction of the beam energy deposited in the ECAL  accompanied by 
 a single muon outgoing from the target and passing  the three HCAL modules, as shown in Fig.\ref{setup}.
 In these searches a sample of $\simeq 10^4$ rare dimuon events from the QED production in the target,  dominated by the  hard bremsstrahlung photon conversion into the $\mu^+ \mu^-$ pair on a target nucleus, $\emu$ was accumulated. 
 Differently from the reaction \eqref{e-react} shown in Fig.\ref{setup},  these events are accompanied by two muons in the final state passing though the HCAL modules.
 They exhibit themselves as a narrow strip  in the measured  distribution of events in the ($E_{ECAL}$;$E_{HCAL}$) plane 
   corresponding to the double MIP (minimum ionizing particle) HCAL  energy  $E_{HCAL}\simeq 12$ GeV  \cite{na64prd18, na64prl19}, see, e.g. Fig. 2 (left panel)  in Ref.\cite{na64prl19} (region I). 
 Using these samples we  define the signal region for events from (\ref{e-react})  to be ($E_{ECAL} < 50$ GeV;  $E_{HCAL} \simeq 6$ GeV) 
where the first cut  is on the missing energy in the ECAL  carried away by the $\varphi$ and the muon,  also used in Ref. \cite{na64prl19} for the search for invisible decays of $A'$s  \cite{na64prd18, na64prl19}); while the second requirement is for the  total energy in  three HCAL modules to be equal  the MIP energy deposited by a single muon.
 \par  Interestingly, several events are observed  in the signal region, the origin of which  is the subject of further detailed analysis beyond the scope of this work. Conservatively attributing these events to  background, we estimated the NA64 sensitivity with  a generic DM simulation package DMG4 \cite{dmg4} used for the signal yield, the efficiency of the signal muon detection and detector acceptance calculations, e.g.,  as in Ref.\cite{na64prl21}.
\begin{figure}[tbh!!]
\begin{center}
\includegraphics[width=0.45\textwidth,height=0.4\textwidth]{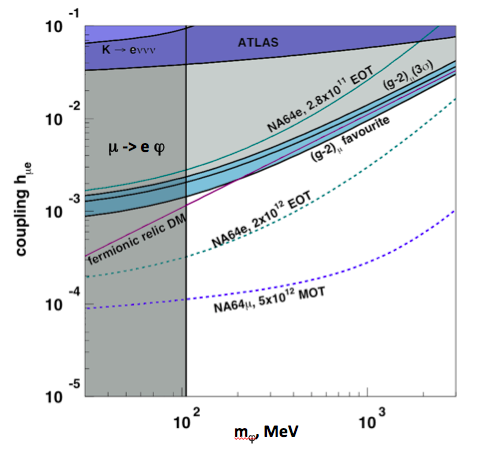}
\caption {The current constraints on the coupling $h_{\mu e}$ in the ($m_\varphi , h_{\mu e}$)  plane (dark dashed area)   from the $\mu \to e \varphi $ \cite{pdg}, $K\to e \nu \nu \nu$ decays \cite{k-e3nu}, the ATLAS experiment \cite{atlas}, and $(g-2)_\mu$ anomaly (3$\sigma$ level). The  90\% C.L.  exclusion regions   from  NA64e with 
$\simeq 2.8\times 10^{11}$ EOT and projection  from NA64e with $\simeq 2\times 10^{12}$ EOT  and  NA64$\mu$ with $\simeq 5\times 10^{12}$ MOT \cite{gkm, laura} are also shown (areas above dashed curves). The $(g-2)_\mu$ favored parameter region (dashed blue area) and the curve explaining the DM relic abundance for fermionic case calculated with   $g_{\varphi \psi_1\psi_2} = 0.08$  assuming $\frac{m_\psi}{m_\varphi} = \frac{1}{3}$ are also shown.
 \label{fig:limits}}
\end{center}
\end{figure} 
\par The combined 90\% C.L. exclusion limits on the coupling parameter $h_{\mu e}$   as a function of the $\varphi$ mass. 
are shown in Fig.~\ref{fig:limits}.  For the region $m_\varphi \lesssim 0.5$ GeV,   NA64  bounds  are more stringent than those derived  from  the  $(g-2)_\mu$
and ATLAS experiment, excluding part of the parameter space favored by the muon anomaly.  
\par  For further searches,  NA64e is planned to  be upgraded with a magnetic spectrometer downstream the HCAL  for the 
measuring of both,  the outgoing muon momentum and, in combination with the ECAL, the missing energy carried away by the $\varphi$, thus allowing significantly improve the search sensitivity. 
\par Another complementary search could be performed  with  the NA64$\mu$  experiment  at  M2 muon beam of the CERN SPS \cite{gkm, laura}
by using the $\varphi$ production in the inverse reaction 
\begin{equation}
\mu Z \to e Z \varphi ; \varphi \to invisible 
\label{mu-react}
\end{equation}
of 100-160 GeV muon scattering on heavy nuclei.
The projection sensitivity for the  $\varphi$ searches with reactions  \eqref{e-react} and \eqref{mu-react} is  shown in Fig.\ref{fig:limits} for the  background-free case.
One can see that with the statistics increased by an order of magnitude one can decisively probe the parameter space explaining the  $(g-2)_\mu$ and the current density of dark matter.  The ($m_\varphi , h_{\mu e}$)  region of interest could also be  effectively tested with the planned M$^3$\cite{krnjaic}  and LDMX  \cite{ldmx,ldmx1,berlin} experiments 
by using the missing momentum technique.
 \par Finally, note that the model additionally  predicts contribution to  the anomalous electron magnetic model at the level  $\Delta a_e = 0.6 \times 10^{-13}$. 
This value is a factor five less then the current error  on $\Delta a_e = (4.8 \pm 3.0) \times 10^{-13}$ determined 
from the recent precise measurements  of the fine-structure constant \cite{lkb}, and hopefully can be probed in the near future.
\par We are grateful to  our colleagues from 
the NA64 Collaboration for their interest, useful discussions,   and  valuable comments. We would also 
like to thank A.N. Toropin for his help in handling the data sample and D.V. Kirpichnikov for the discussion on limit calculations.


\begin{thebibliography}{99}

\bibitem{fnal}
B. Abi  {\it et al.}, Phys. Rev. Lett. {\bf 126}, 141801 (2021)

\bibitem{bnl}
 G.W. Bennett  {\it et al.}, Phys. Rev. D {\bf 43}, 072003 (2006).

\bibitem{th1}
T.  Aoyama  {\it et al.}, Phys. Rep. {\bf 887}, 1 (2020).

\bibitem{th2}
T. Aoyama, M. Hayakawa, T. Kinoshita,  and M. Nio, Phys. Rev. Lett. {\bf 109}, 111808 (2012).

\bibitem{th3}
 K. Melnikov and A. Vainshtein, Phys. Rev. D {\bf 70}, 113006 (2004).

\bibitem{th4}
A. Czarnecki, W. J. Marciano,  and A. Vainshtein,
Phys. Rev. D {\bf 67}, 073006 (2003); [erratum: Phys. Rev. D {\bf 73}, 119901 (2006)].

\bibitem{th5}
A. Kurz, T. Liu, P. Marquard,  and M. Steinhauser, Phys. Lett. B {\bf 734}, 144 (2014).

\bibitem{th6}
M. Hoferichter, B. L. Hoid, B. Kubis, S. Leupold,  and S. P. Schneider, JHEP {\bf 10}, 141 (2018).


\bibitem{th7}
A. Keshavarzi, D. Nomura, and T. Teubner, Phys. Rev. D {\bf 97}, 114025 (2018);
Phys. Rev. D {\bf 101}, 014029 (2020).

\bibitem{thh8}
G. Colangelo, M. Hoferichter,  and P. Stoffer, JHEP {\bf 02}, 006 (2019).

\bibitem{thh9}
M. Hoferichter, B. L. Hoid,  and B. Kubis, JHEP {\bf 08}, 137 (2019). 

\bibitem{thh10}
A. Gerardin, H. B. Meyer,  and A. Nyffeler, Phys. Rev. D {\bf 100}, 034520 (2019).

\bibitem{thh11}
M. Davier, A. Hoecker, B. Malaescu, and Z. Zhang, Eur. Phys. J. C {\bf 80}, 241 (2020); [erratum: Eur. Phys. J. C
{\bf 80},  410 (2020)]. 


\bibitem{thh12}
T. Blum, N. Christ, M. Hayakawa, T. Izubuchi, L. Jin, C. Jung,  and C. Lehner, Phys. Rev. Lett. {\bf 124},132002 (2020).

\bibitem{g-2_np}
P. Athron, C. Bal\'azs, D.H. Jacob, W. Kotlarski, D. St\"ockinger and H. St\"ockinger-Kim, JHEP {\bf 09}, 080 (2021). 

\bibitem{gk1}
  S.~N.~Gninenko and N.~V.~Krasnikov,  Phys.\ Lett.\ B {\bf 513}, 119 (2001).
  
\bibitem{gkm}
 S.~N.~Gninenko, N.~V.~Krasnikov, and V.~A.~Matveev, Phys. Rev. D {\bf 91},  095015 (2015).  
  
 \bibitem{gk2}
  S.~N.~Gninenko and N.~V.~Krasnikov, Phys.\ Lett.\ B {\bf 783}, 24 (2018).
     
\bibitem{laura}  
H. Sieber, D. Banerjee, P. Crivelli, E. Depero, S. N. Gninenko, D. V. Kirpichnikov, M. M. Kirsanov, V. Poliakov, and L. Molina Bueno,  Phys. Rev. D. {\bf 105},  052006 (2022).

\bibitem{laurakirp}
D. V. Kirpichnikov, H. Sieber, L. Molina Bueno, P. Crivelli, and M. M. Kirsanov, Phys. Rev. D {\bf 104}, 076012 (2021).   

\bibitem{Kirpichnikov:2020tcf}
  D.~V.~Kirpichnikov, V.~E.~Lyubovitskij, and A.~S.~Zhevlakov,
  Phys.\ Rev.\ D {\bf 102},  095024 (2020).  
\bibitem{na64epjc21}
C. Cazzaniga {\it et al.} (NA64 Collaboration),   Eur. Phys. J. C {\bf 81}, 959 (2021).

\bibitem{krnjaic} 
Y. Kahn, G. Krnjaic, N. Tran, and A. Whitbeck,  JHEP {\bf 09}, 153 (2018).

\bibitem{krnjaic1}
I. Holst, D. Hooper, and G. Krnjaic, Phys. Rev. Lett. {\bf 128}, 141802 (2022).

\bibitem{krnjaic2}
R. Capdevilla, D. Curtin, Y. Kahn, and G. Krnjaic, arXiv:2112.08377.


\bibitem{chen}
Chien-Yi Chen, M. Pospelov, and Yi-Ming Zhong,  Phys. Rev. D {\bf 95}, 115005 (2017).

\bibitem{jr}
  J.~Jaeckel and A.~Ringwald,
   Annu.\ Rev.\ Nucl.\ Part.\ Sci.\  {\bf 60}, 405 (2010).

\bibitem{Essig:2013lka} 
  R.~Essig {\it et al.},  arXiv:1311.0029.


 \bibitem{report1}
  J.~Alexander {\it et al.}, arXiv:1608.08632.

\bibitem{report2} 
  M.~Battaglieri {\it et al.},  arXiv:1707.04591.


\bibitem{pbc-bsm}
J.  Beacham  {\it et al.},  J.\ Phys.\ G {\bf 47},  010501 (2020); arXiv:1901.09966.   


\bibitem{dama} H.Davvoudiasl and W.J.Marciano, Phys. Rev. D {\bf 98}, 075011 (2018) .

\bibitem{cdmz}  C. Y. Chen, H. Davoudiasl, W. J. Marciano, and C. Zhang, Phys. Rev. D {\bf 93}, 035006 (2016).

\bibitem{bbat}  B. Batell, N. Lange, D. McKeen, M. Pospelov, and A. Ritz, Phys. Rev. D {\bf 95}, 075003 (2017).

\bibitem{gk161} S.N. Gninenko and N.V. Krasnikov, EPJ Web Conf. {\bf 125}, 02001 (2016).

\bibitem{gk162}  S.N. Gninenko and N.V. Krasnikov, Mod. Phys. Lett. A  {\bf 31}, 1650142 (2016).  

\bibitem{pdg} 
P.~A.~Zyla {\it et al.} (Particle Data Group), Prog. Theor. Exp. Phys. {\bf 2020}, 083C01 (2020).

\bibitem{muonmoment} As a review, see for example,  
F. Jegerlehner and A. Nyffeler,  Phys. Rep. {\bf 477} 1 (2009).  

\bibitem{Planck} P.A.R. Ade  {\it et al.} (Planck Collaboration),   Astron. Astrophys.  A {\bf 13} 594 (2016).

\bibitem{c1} E.W. Kolb and M. S.  Turner,  Front. Phys.{\bf 69} 1  (1990).

\bibitem{c2}  D.S. Gorbunov and V.A.  Rubakov, {\it Introduction to the theory 
of the early Universe} (World Scientific Publishing Co. Pt. Ltd.,  Singapore, 2011). 

\bibitem{c3} P. Gondolo and G. Gelmini,  Nucl. Phys.  {\bf B360} 145 (1991). 

\bibitem{c4} S. N. Gninenko, N. V.  Krasnikov, and V. A.  Matveev,  Phys. Part. Nucl.  {\bf 51} 829 (2020).

\bibitem{c5} S. N. Gninenko, N. V.  Krasnikov, and V. A.  Matveev,   Usp. Fiz. Nauk {\bf 191} 1361 (2021).

\bibitem{sp}
 S. Profumo, arXiv:1301.0952.

\bibitem{atlas}
G. Aad  {\it et al.} (ATLAS Collaboration),  Phys. Rev. D {\bf 90}, 072010 (2014).

\bibitem{k-e3nu}  
J. Heintze {\it et al.} , Nucl. Phys.  {\bf B149}, 365 (1979). 



 \bibitem{na64prl17} 
  D.~Banerjee {\it et al.} (NA64 Collaboration),
  Phys.\ Rev.\ Lett.\  {\bf 118},   011802 (2017).
 

\bibitem{na64prd18} 
  D.~Banerjee {\it et al.} (NA64 Collaboration),
  Phys.\ Rev.\ D {\bf 97},  072002 (2018).


 \bibitem{na64prl19} 
  D.~Banerjee {\it et al.} (NA64 Collaboration),
  Phys.\ Rev.\ Lett.\  {\bf 123},  121801 (2019).

\bibitem{Banerjee:2015eno}
  D.~Banerjee, P.~Crivelli, and A.~Rubbia,
   Adv.\ High Energy Phys.\  {\bf 2015},  105730 (2015).


\bibitem{na64srd}  
  E.~Depero {\it et al.},
  Nucl.\ Instrum.\ Methods\ Phys. Res., Sect. A {\bf 866},  196 (2017).

\bibitem{dmg4} 
A. Celentano, M. Bondi, R. R. Dusaev,  D. V. Kirpichnikov, M. M. Kirsanov, N. V. Krasnikov, L. Marsicano, and D. Shchukin, 
    Comput. Phys. Commun. {\bf 269},   108129 (2021). 
    

 \bibitem{na64prl21} 
  Yu.M..~Andreev {\it et al.} (NA64 Collaboration),
  Phys.\ Rev.\ Lett.\  {\bf 126},   211802 (2021).
    
\bibitem{ldmx}
 T. Akesson  {\it et al.}  (LDMX Collaboration), arXiv:1808.05219.

\bibitem{ldmx1}
 T. Akesson  {\it et al.}  (LDMX Collaboration),   JHEP {\bf 04}, 003 (2020).
 
\bibitem{berlin}
A. Berlin, N. Blinov, G. Krnjaic, P. Schuster, and N. Toro,
Phys. Rev. D {\bf 99}, 075001 (2019).

 
\bibitem{lkb}
L.~Morel, Zh.~Yao, P.~Clad\'e,  and S.~Guellati-Kh\'elifa, Nature (London) {\bf 588}, 61 (2020).



 
\end{thebibliography}
\end{document}